 \documentclass[preprint,aps,amsmath,amssymb,showpacs]{revtex4}

\usepackage{graphicx}% Include figure files
\usepackage{dcolumn}% Align table columns on decimal point
\usepackage{bm}% bold math
\usepackage{amssymb} %extra symbol
\usepackage[usenames]{color}

\begin{document}

\title{The Electron Glass in a Switchable Mirror: Relaxation, Aging and Universality }

\author{M. Lee$^{1}$, P. Oikonomou$^{1}$, P. Segalova$^{1}$, T.F. Rosenbaum$^{1}$, A.F.Th. Hoekstra$^{2}$ and P.B. Littlewood$^{3}$}

\affiliation{$^{1}$James Franck Institute and Department of Physics,
 The University of Chicago, Chicago, Illinois 60637\\
 $^{2}$Kammerlingh Onnes Laboratory, Leiden University, 2300 RA Leiden, The Netherlands\\
$^{3}$Cavendish Laboratory, University of Cambridge, Cambridge, CB3 0HE, UK}

\date{\today}

\begin{abstract}
    
The rare earth hydride YH$_{3-\delta}$ can be tuned through the metal-insulator transition both by changing $\delta$ and by illumination with ultraviolet light. The transition is dominated by strong electron-electron interactions, with transport in the insulator sensitive to both a Coulomb gap and persistent quantum fluctuations. Via a systematic variation of UV illumination time, photon flux, separation between electrons, and temperature, we demonstrate that polycrystalline  YH$_{3-\delta}$ serves as a model system for studying the properties of the interacting electron glass.  Prominent among its features are logarithmic relaxation, aging, and universal scaling of the conductivity.
\end{abstract}

\maketitle

Glassy systems can be trapped in local minima in the free energy surface, unable to fully explore phase space. They are then characterized by long-time relaxation of non-exponential form and memory of their particular path of descent through the free energy surface. Materials as diverse as supercooled liquids, disordered and frustrated magnets, and sliding charge-density waves all share these signatures of glassiness, although the microscopic underpinnings of their behavior still remain a challenge to understand. A glass composed of randomly distributed electrons, point charges rather than complex structural units, should provide a simpler and perhaps more tractable model for glassy behavior.

The electron glass has been considered in both theory and simulation \cite {Pollak, Mott, Davies, Yu}, with an emphasis on the many-body effects that should contribute to the long-time relaxation. The long-range nature of the Coulomb interaction can give rise to complex rearrangements of electrons across a spectrum of free energies and times.  Low temperature transport in doped semiconductors \cite{Pollakbook, Paalanen, marklee} as well as the decay of injected carriers in MOSFETs \cite{Monroe} have been interpreted in these terms. Recently, in a series of definitive experiments \cite{Ovadyahu}, Ovadyahu and colleagues have studied the non-ergodic transport properties of $\alpha$-InO$_x$, an amorphous alloy oxidized into the insulating state. By adding charge through a gate electrode and carefully correlating the evolution of the conductivity with the injection history, they were able to chronicle both a logarithmically slow relaxation and the system's memory of previous excitations. 
	
We report here electron glass behavior in polycrystalline yttrium hydride, a correlated electron system whose insulating state manifests a Coulomb gap as well as strong quantum fluctuations in the near vicinity of the Mott-Hubbard metal-insulator transition. Dynamical scaling is consistent solely with the emergence of a Coulomb gap which collapses rapidly at the approach to the transition from below \cite{Arunabha}.  Charge carriers can be introduced via exposure to UV light, and the resulting long-time relaxation of the conductivity can be collapsed onto a universal curve for different illumination times, photon fluxes, and doping levels. The relaxation of the photo-induced conductivity converges to a generic logarithmic function of time, but only after a lapse of time corresponding to the historical illumination period.  The weak temperature dependence of the relaxation points to the pivotal role played by collective rearrangements of correlated electrons.
 
%%%%%%%%%% Figure1	 begins %%%%%%%%%%
\begin{figure}
\begin{center}
\includegraphics[width=3.25 in]{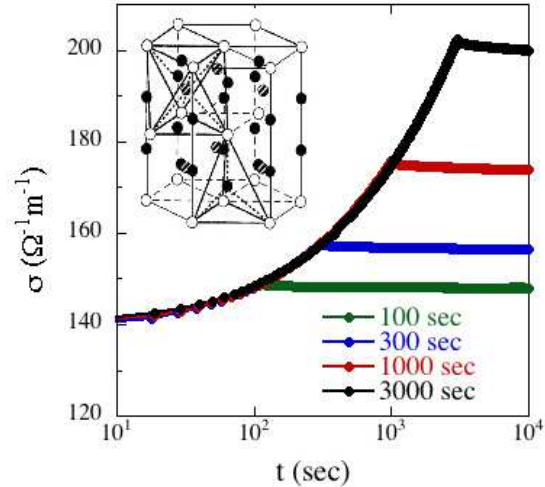}
\caption{
Conductivity change as a function of time after four illuminations for a polycrystalline YH$_{3-\delta}$ sample (structure in inset, following Ref. \cite {Huiberts_thesis}. Open circles indicate Y sites and the striped and solid circles are H atoms at octahedral and tetrahedral sites, respectively). A sharp increase in conductivity follows immediately after the UV light is turned on at $t = 0$ and relaxation starts after turning off the lamp. Each curve is obtained at $T = 4.2$ K after resetting the system by raising its temperature above 250 K (where hydrogen vacancies are free to move). The curves overlap, indicating that there is no significant change caused by thermal cycling. 
}
\end{center}
\end{figure}
%%%%%%%%%% Figure1 Ends %%%%%%%%%%%%%

The rare earth hydrides first attracted attention as Òswitchable mirrors.Ó They can be transformed rapidly from metal to insulator, from reflective mirror to transparent window, simply by changing the surrounding hydrogen gas pressure \cite {TFR_Allard}. 
Tuning the quantum phase transition by light at sub-Kelvin temperatures in YH$_{3-\delta}$ established that the metal-insulator transition is in the highly correlated limit, with electron-electron interactions playing an important role even deep in the insulator. Exposure to UV light excites the system away from equilibrium, enhancing its conductivity, which then slowly relaxes towards its new equilibrium state when the light is turned off. The new charge carriers persist, presumably resulting from changes in the bonding configurations between the yttrium and the hydrogen \cite {Roy_thesis}, but adjust themselves to their new environment over periods of at least $10^6$ sec. It is the systematics of this adjustment in a microscopically well-defined and controlled crystalline material with random placement of electrons that is the focus of our studies.

%%%%%%%%%% Figure2	 begins %%%%%%%%%%
\begin{figure*}
\begin{center}
\includegraphics[width=7 in]{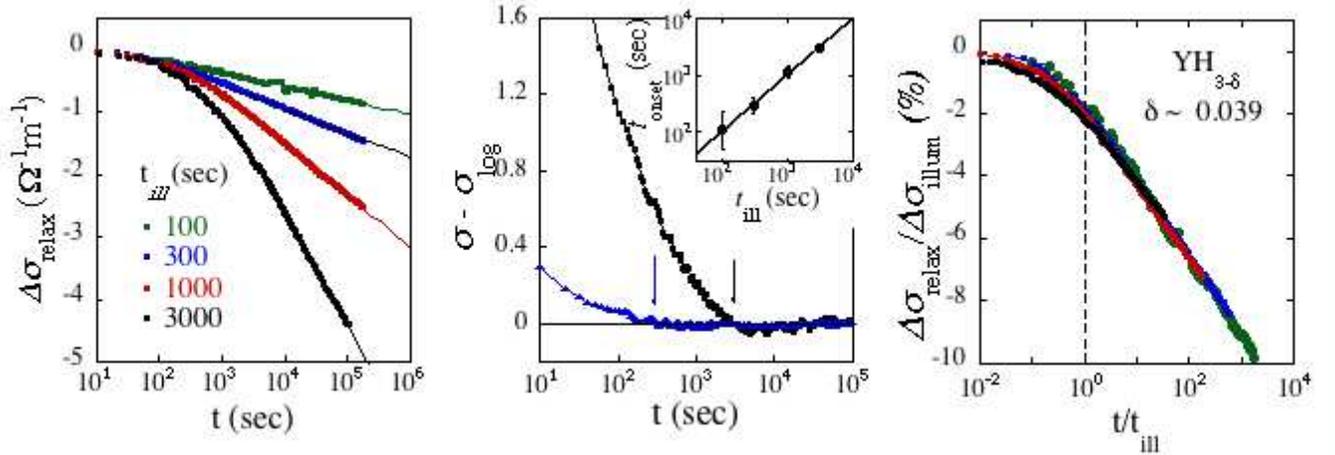}
\caption{
The relaxation after different illumination times at fixed temperature, $T = 4.2$ K. All data were taken for $\delta  \sim 0.039$ (most insulating hydrogenation). (a) The change in the conductivity, $\Delta\sigma_{\mathrm{relax}} (t) = \sigma (t)- \sigma(0)$, is greater and faster for longer illumination times. 
(b) The relaxation becomes logarithmic after a decay time equal to the original illumination time, $t_{\mathrm{ill}}$. This can be seen by plotting the point designated by the arrows at which $\log (t)$ behavior starts, $t_{\mathrm{onset}}$ vs. $t_{\mathrm{ill}}$ (inset).
(c)  All curves are collapsed onto a universal curve when $\Delta\sigma_{\mathrm{relax}} (t)$ is normalized by the conductivity gain from the illumination, $\Delta\sigma_{\mathrm{illum}} = \sigma (t=0)-\sigma_0$, where $\sigma_0$ is the conductivity before UV exposure and $t$ is normalized by $t_{\mathrm{ill}}$. The vertical line specifies $t/t_{\mathrm{ill}} =$ 1.
 }
\end{center}
\end{figure*}
%%%%%%%%%% Figure2 Ends %%%%%%%%%%%%%
A 550 nm thick film of Y was covered with 5 nm of Pd to allow hydrogen diffusion and prevent oxidation. The Pd layer is segmented into $\sim$10 nm wide disconnected islands separated by Y$_2$O$_3$ to prevent electrical shorts. The film is mounted in a specially designed hydrogen loading cell with electrical and optical access \cite{Allard_cell} in a $^3$He cryostat. Light from an ultraviolet stroboscope (spectral range 220-700 nm, maximum repetition rate 240 Hz) is guided to the cell by a UV silica fiber that illuminates the entire sample area. Four copper wires are connected at the perimeter of the 7 mm diameter disk-shaped film to determine its conductivity using the van der Pauw configuration and a standard lock-in technique. The data are in the Ohmic and frequency-independent regimes for all $T$ and $\delta$.

The typical experimental procedure is illustrated in Fig. 1. The conductivity $\sigma$ of YH$_{3-\delta}$ is monitored as a function of time $t$ at the measuring temperature, $T_m$. The distance below the insulator-metal transition at $\delta = 0$.15 can be controlled by the hydrogenation pressure, p(H$_2$) at room temperature. For the crystal in Fig. 1, p(H$_2$) was 140 bar, corresponding to $\delta \sim 0.039$. After cooling to 4.2 K, we wait 7 to 10 hours to assure thermal equilibration, and then expose the sample to ultraviolet light. During illumination for time $t_{\mathrm{ill}}$, the time the stroboscope is on, $\sigma$ increases sharply while the temperature is stabilized at 4.2 K within a few mK. The enhanced $\sigma$ due to the UV illumination can only be restored to its initial value by raising the system temperature above $T \sim 250$ K, at which point the hydrogen vacancies become mobile \cite {Huiberts97}. The four curves in Fig. 1 correspond to four different illumination times, with each curve obtained after warming to room temperature to reset the system. The overlap of the curves before relaxation demonstrates that the initial conditions are identical in each case. 
%%%%%%%%%% Figure3	 begins %%%%%%%%%%
\begin{figure}
\begin{center}
\includegraphics[width=3.2 in]{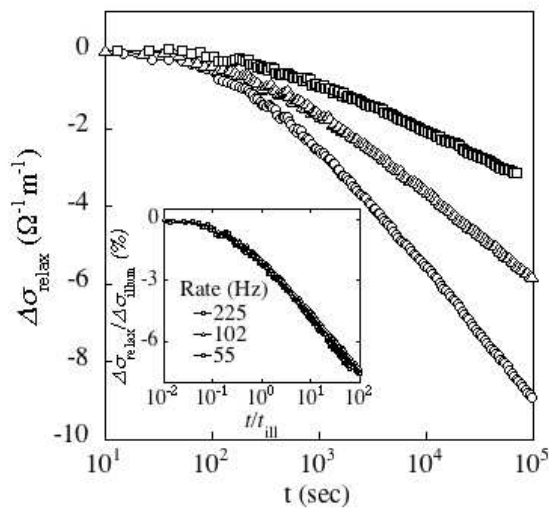}
\caption{
The relaxation of the YH$_{3-\delta}$ sample with $\delta \sim 0.062$ after 1000 sec illumination at $T = 77$ K with repetition rates of the pulsed UV lamp from 55 to 225 Hz. Collapse of the curves in the inset proves that it is indeed illumination time $t_{\mathrm{ill}}$ rather than the total number of photons (repetition rate) that is the proper scaling variable. 
}
\end{center}
\end{figure}
%%%%%%%%%% Figure 3 Ends %%%%%%%%%%%%%
%%%%%%%%%% Figure 4	 Begins %%%%%%%%%%
\begin{figure}
\begin{center}
\includegraphics[width=3.2 in]{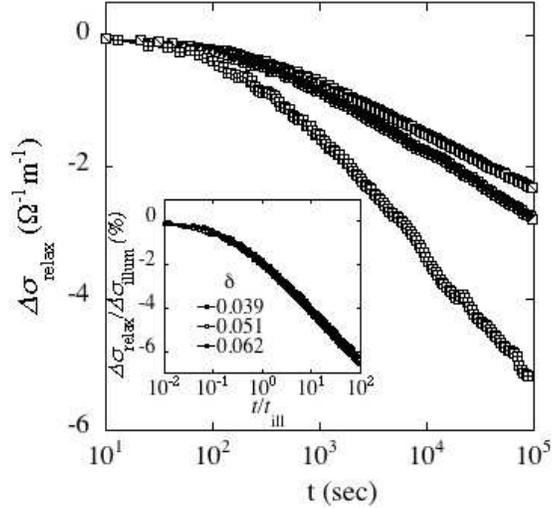}
\caption{
The relaxation after 1000 sec illumination at $T = 4.2$ K for different depths of the Coulomb gap, regulated by hydrogen level. Universal behavior is found using the same scaling form (inset). 
}
\end{center}
\end{figure}
%%%%%%%%%% Figure 4 Ends %%%%%%%%%%%%%

After turning off the lamp, $\sigma$ starts to decrease as the glass attempts to approach its new equilibrium condition. We now concentrate on the nature of this relaxation and define $t = 0$ as the time at which the lamp is switched off. The change in the conductivity, $\Delta\sigma_{\mathrm{relax}}(t) = \sigma (t) - \sigma (0)$, is shown for a set of different illumination times in Fig. 2a. Both the shape and the magnitude of the relaxation depend on $t_{\mathrm{ill}}$. In the long time limit, $\Delta\sigma_{\mathrm{relax}} (t)$ follows a $\log (t)$ form with different slopes for different $t_{\mathrm{ill}}$. We plot in Fig. 2b the difference between $\Delta\sigma_{\mathrm{relax}}(t)$ and the best-fit $\log(t)$ curve, where $t_{\mathrm{onset}}$ is defined as the time at which $\Delta\sigma_{\mathrm{relax}}(t)$ crosses over to its long-time logarithmic behavior. This crossover time corresponds precisely to $t_{\mathrm{ill}}$ (Fig. 2b inset, with best fit slope of $0.99 \pm 0.03$ on a $\log$-$\log$ scale). The YH$_{3-\delta}$ crystal is imprinted with the history of its previous condition (lamp-on) even after the condition has been removed (lamp-off), displaying the fundamental quality of a glassy system known as ``aging".

It is possible to collapse all the curves onto a universal function by normalizing the abscissa by $t_{\mathrm{ill}}$ and the ordinate by the excess conductivity, $\Delta\sigma_{\mathrm{illum}} = \sigma (t=0) - \sigma_0$, where $\sigma_0$ is the conductivity before illumination (Fig. 2c). The need to normalize by the historical excess conductivity derives from the persistent nature of the photoconductivity peculiar to the switchable mirrors. This scaling demonstrates that the slow dynamics are correspondingly due to electronic relaxation rather than structural glassiness. By contrast, the logarithmic time dependence is a generic glassy feature, observed for a variety of systems with a broad distribution of barriers to relaxation, including amorphous thin metal films 
\cite{Ovadyahu, Goldman98}, semiconductors \cite {Queisser86} and spin glasses \cite {Norblad}. 

Given that the UV light source used in this experiment is pulsed, it is fair to ask whether the universal relaxation actually depends on the total time of illumination, $t_{\mathrm{ill}}$, or the total incident flux (i.e. the number of photons). In order to clarify this issue, we performed a series of measurements with a fixed illumination time, $t_{\mathrm{ill}} = 1000$ sec, and a fixed pulse length of 5 $\mu$sec, but different pulse repetition frequencies. If a small repetition rate is simply equivalent to a shorter total illumination time, then $t_{\mathrm{onset}}$ will no longer coincide with $t_{\mathrm{ill}}$. We find in Fig. 3, however, that the pertinent scaling quantity is indeed $t_{\mathrm{ill}}$: the onset of logarithmic decay is determined by $t_{\mathrm{ill}}$ regardless of the pulse frequency. The increased dosage of photons with increased repetition rate does create more carriers and, consequently, a larger relaxation of the conductivity, but when properly normalized by $\Delta\sigma_{\mathrm{illum}}$ as before, the data again collapse onto a master curve (Fig. 3, inset). 
There remain two fundamental parameters to vary that can test the simple scaling relationship that we have uncovered: the strength of the interaction as modulated by the reciprocal separation between electrons and temperature. We show in Fig. 4 the result of changing the degree of hydrogenation. Smaller values of $\delta$ correspond to increasingly insulating behavior and a deeper Coulomb gap, with the average separation between electrons decreasing by approximately 10$\%$ for $\delta$ ranging from 0.039 to 0.062 \cite {comment_T0}. The conductivity gain due to illumination as well as the magnitude of the relaxation decrease as the system is tuned deeper into the insulator, but taking their ratio recovers the usual scaling behavior (Fig. 4, inset). In all cases, the onset of relaxation is found at $t = t_{\mathrm{ill}} = 1000$ sec.

Finally, we turn to the effects of temperature. We plot in Fig. 5 the relaxation of the conductivity after 1000 sec illumination over 2.5 decades of temperature, $0.4 < T < 140$ K. As expected, increasing T increases $\Delta\sigma_{\mathrm{relax}}$, but the effect is small and certainly not exponentially dependent on temperature as would be expected for simple activation over a single free energy barrier \cite {Grenet03}.  Collapse of all the temperature data is almost possible (Fig. 5, inset), but certainly not up to the standards of Figs. 2-4.  The obvious exception is the $T = 142$ K response where curvature on a $\log(t)$ scale is discernible and hydrogen vacancy movement may start to enter the physics.  We have checked that illumination at room temperature yields marginal changes in the conductivity ($< 0.4\%$), with no indication of a logarithmic time dependence.

The subordinate role played by temperature provides evidence for the many-body nature of the electron glass.  Multiple barriers to relaxation, growing larger as correlated electrons execute increasingly more complex rearrangements, can guarantee that there are always free energy barriers much larger than the temperatures accessed in this work.  Moreover, a hierarchical distribution of barrier potentials \cite {Palmer84}, with electrons running into higher and higher barriers as relaxation proceeds, leads generically to a long-time relaxation of logarithmic form \cite {Sibani}. This $\log(t)$ behavior is independent of the details of the barrier distribution and, in fact, is found in our experiments to be universal under a wide variety of perturbations to the free energy surface. 
%%%%%%%%%% Figure 5	 begins %%%%%%%%%%
\begin{figure}
\begin{center}
\includegraphics[width=3.2 in]{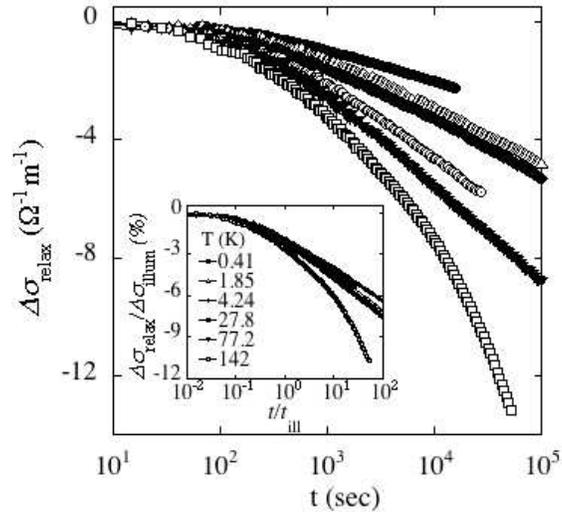}
\caption{
Variation of the relaxation with temperature after 1000 sec illumination. As $T$ decreases, the magnitude of the relaxation also decreases but only weakly, pointing to the many-body nature of the electron glass (see text).
}
\end{center}
\end{figure}
%%%%%%%%%% Figure 5 Ends %%%%%%%%%%%%%

We are indebted to R. Griessen for introducing us to the fascinating properties of the rare earth hydrides. The work at the University of Chicago was supported by the National Science Foundation under Grant No. DMR-0114798.

\end{document}